\documentclass[apj]{emulateapj}
\usepackage{apjfonts}

\gdef\gala{FW-4871}
\gdef\galb{FW-4887}
\gdef\kms{km\,s$^{-1}$}
\gdef\msun{${\rm M}_{\odot}$}
\lefthead{van Dokkum \& Brammer}
\righthead{A Growing Compact Galaxy at $z=2$}
\slugcomment{draft, \today}
\begin{document}

\title{
Hubble Space Telescope WFC3 Grism Spectroscopy and Imaging of a Growing
Compact Galaxy at $z=1.9$}

\author{Pieter G.\ van Dokkum
and
Gabriel Brammer
}

\affil
{Department of Astronomy, Yale University, New Haven, CT 06520-8101}

\begin{abstract}
We present HST/WFC3 grism spectroscopy of the brightest
galaxy at $z>1.5$ in the
GOODS-South WFC3 Early Release Science grism pointing, covering
the  wavelength range $0.9\,\mu$m -- 1.7\,$\mu$m. The spectrum
is of remarkable quality
and shows the redshifted Balmer lines H$\beta$, H$\gamma$, and H$\delta$
in absorption at $z=1.902 \pm 0.002$, correcting previous erroneous
redshift measurements from the rest-frame UV. The 
average rest-frame equivalent width
of the Balmer lines is 8\,\AA\,$\pm 1$\,\AA, which can be
produced by a post-starburst
stellar population with a luminosity-weighted age
of $\approx 0.5$\,Gyr.
The $M/L$ ratio inferred from the spectrum
implies a stellar mass
of $(4\pm 1) \times 10^{11}$\,\msun. We determine the morphology of the
galaxy from a deep WFC3 $H_{160}$ image.
Similar to other massive galaxies
at $z\sim 2$ the galaxy is compact, with
an effective radius of $2.1\pm 0.3$\,kpc.
Although most of the light is in a compact core, the galaxy has
two red, smooth spiral arms that appear to be tidally-induced.
The spatially-resolved spectroscopy demonstrates
that the center of the galaxy is quiescent and the surrounding
disk is forming stars, as it shows H$\beta$ in emission.
The galaxy is interacting with a companion at a projected
distance of 18\,kpc,
which also shows prominent tidal features.
The companion has a slightly redder spectrum than
the primary galaxy but is a factor of $\sim 10$ fainter and
may have a lower metallicity.
It is tempting to interpret these
observations as ``smoking gun'' evidence for the growth of compact,
quiescent high redshift galaxies through minor mergers, which has
been proposed by several recent observational and theoretical studies.
Interestingly both objects host luminous AGNs, as indicated
by their X-ray luminosities, which implies 
that these mergers can be accompanied by
significant black hole growth. This study illustrates the
power of moderate dispersion, low background near-IR
spectroscopy at HST resolution, which is now available with
the WFC3 grism. 
\end{abstract}


\section{Introduction}

\noindent
The formation history of massive galaxies is not well understood.
Present-day galaxies with stellar masses $\gtrsim 3\times 10^{11}$\,\msun\
are typically giant elliptical galaxies in the centers of galaxy groups.
These galaxies have old stellar populations and follow tight scaling
relations between their velocity dispersions,
sizes, surface brightnesses, line strengths, and other parameters
(e.g., {Djorgovski} \& {Davis} 1987; {Thomas} {et~al.} 2005).
At redshifts $z\sim 2$ massive galaxies form a more complex
population. A fraction of the population is forming stars at a high
rate, as determined from their brightness in the rest-frame UV or IR,
emission lines such as H$\alpha$, and other indicators 
(e.g., {Steidel} {et~al.} 1996; {Blain} {et~al.} 2002; {Rubin} {et~al.} 2004; {Papovich} {et~al.} 2006, and many other
studies). However,
others have no clear indications of ongoing star formation and have
spectral
energy distributions (SEDs) characterized by
strong Balmer- or 4000\,\AA\ breaks
(e.g., {Daddi} {et~al.} 2005;
{Kriek} {et~al.} 2006).
The existence of these ``quiescent'' galaxies at this early epoch is in
itself remarkable, and provides constraints on the
accretion and thermodynamics of gas
in massive halos at $z>2$
(e.g., Kere{\v s} et al.\ 2005; {Dekel} \& {Birnboim} 2006). What is
perhaps even more surprising
is that these galaxies
are structurally very different from early-type galaxies in the
nearby Universe:
their effective radii are typically 1--2\,kpc,
much smaller than nearby giant ellipticals
(e.g., {Daddi} {et~al.} 2005; {Trujillo} {et~al.} 2006;
{van Dokkum} {et~al.} 2008; {Cimatti} {et~al.} 2008).

Several explanations have been offered for the dramatic size
difference between local massive galaxies and quiescent
galaxies at high redshift.
The simplest is that observers underestimated
the sizes and/or overestimated the masses. Although subtle errors are
almost certainy present in the interpretation of the
data, recent studies suggest
that it is difficult to change the sizes and the masses
by more than a factor of 1.5, unless the IMF is altered (e.g.,
Muzzin et al.\ 2009, Cassata et al.\ 2010, Szomoru et al.\ 2010).
Other explanations include 
extreme mass loss due to a quasar-driven wind
({Fan} {et~al.} 2008), strong radial age gradients leading to large
differences between mass-weighted and luminosity-weighted ages
({Hopkins} {et~al.} 2009; {La Barbera} \& {de Carvalho} 2009), star formation due to gas
accretion ({Franx} {et~al.} 2008),
and selection effects
(e.g., {van Dokkum} {et~al.} 2008; {van der Wel} {et~al.} 2009).
Perhaps the most plausible mechanism for
bringing the compact $z\sim 2$ galaxies onto the local
mass-size relation is (minor) merging
(e.g., {Bezanson} {et~al.} 2009; {Naab}, {Johansson}, \& {Ostriker} 2009; {van Dokkum} {et~al.} 2010; Carrasco, Conselice, \& Trujillo 2010).
Numerical
simulations predict that such mergers are frequent
({Guo} \& {White} 2008; {Naab} {et~al.} 2009);
furthermore, they may lead to stronger
size growth than mass growth ({Bezanson} {et~al.} 2009).
From an analysis of mass evolution at fixed number density,
{van Dokkum} {et~al.} (2010) infer that massive galaxies have doubled
their mass since $z=2$, and suggest that $\sim 80$\,\% of
this mass growth can be attributed to mergers.

\setcounter{figure}{1}
\noindent
\begin{figure*}[t]
\epsfxsize=17cm
\epsffile[86 528 563 688]{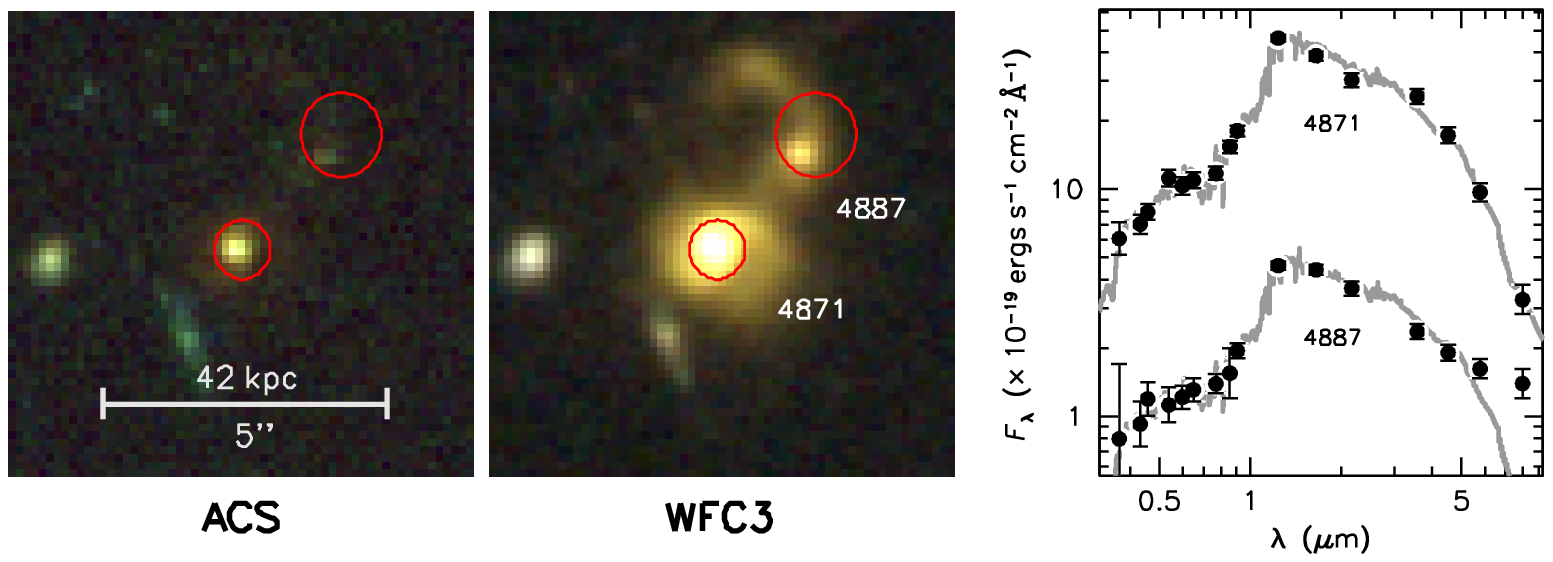}
\caption{\small HST ACS and WFC3 images of \gala\
and its companion \galb. The ACS color image
was created from the $B_{435}$, $V_{606}$, and $z_{850}$ bands, and
the WFC3 image from the $Y_{098}$, $J_{125}$, and $H_{160}$ bands.
\gala\ has a compact core and spiral arms, which may be
the result of an interaction with \galb. Red circles are the
locations of X-ray sources in the Luo et al.\ (2008)
catalog, with the size of the circles indicating the uncertainties
in the positions. Both galaxies host an AGN.
The SEDs of the two galaxies
(from Wuyts et al.\ 2008) are shown in the right-most panel. The
galaxies are both red and have broadly similar SEDs. 
\vspace{0.0cm}
\label{imspec.plot}}
\end{figure*}
Although qualitatively consistent with observations and theory
the minor merger scenario currently has little
direct evidence to support it. It is also not clear whether
properties other than sizes and masses are easily
explained in this context; one of the open questions is why
present-day elliptical galaxies are so red and homogeneous if half
of their mass was accreted from the general field at relatively recent
times. Ideally we would identify and study the infalling population
directly at high redshift, but so far this has been hampered by the
limitations of ground-based spectroscopy and ground- and space-based
near-IR imaging.

In this {\em Letter},
we use the exquisite WFC3 grism on the {\em Hubble Space Telescope}
(HST), in combination with WFC3 imaging, to study the environment of
a quiescent compact galaxy at $z=1.9$. As we show below, the observations
presented here
provide the first direct evidence for minor mergers
as a mechanism for the growth of compact galaxies at high
redshift.
We use $H_0=70$\,\kms\,Mpc$^{-1}$, $\Omega_m=0.3$, and
$\Omega_{\Lambda}=0.7$. Magnitudes are on the AB system.

\section{Selection and Basic Data}

\noindent
The Early Release Science (ERS) WFC3 imaging
observations of the GOODS-South
field comprise a mosaic of eight HST pointings. All eight pointings
were observed with a suite of imaging filters but only one was observed
with the G102 and G141 grisms. The grism data are important for
measuring the redshifts, ages, and star formation rates
of massive galaxies at high redshift and indispensable for measuring the
redshifts of any faint companion galaxies. The G141 grism is
particularly useful as it is very sensitive and its wavelength
range of $1.1\,\mu$m -- $1.7\,\mu$m covers
the redshifted Balmer lines, 4000\,\AA\ break, and
[O\,{\sc iii}] emission at $z\sim 2$. Here we concentrate on
the brightest galaxy
at $z>1.5$ in the $2'\times 2'$ grism field,
indicated with the arrow in Fig.\ \ref{select.plot}.
The galaxy has ID number 4871 in the
$K$-selected FIREWORKS catalog of GOODS-South ({Wuyts} {et~al.} 2008),
and has a total $K$ magnitude of 19.7.

\setcounter{figure}{0}
\noindent
\begin{figure}[htbp]
\epsfxsize=8.5cm
\epsffile{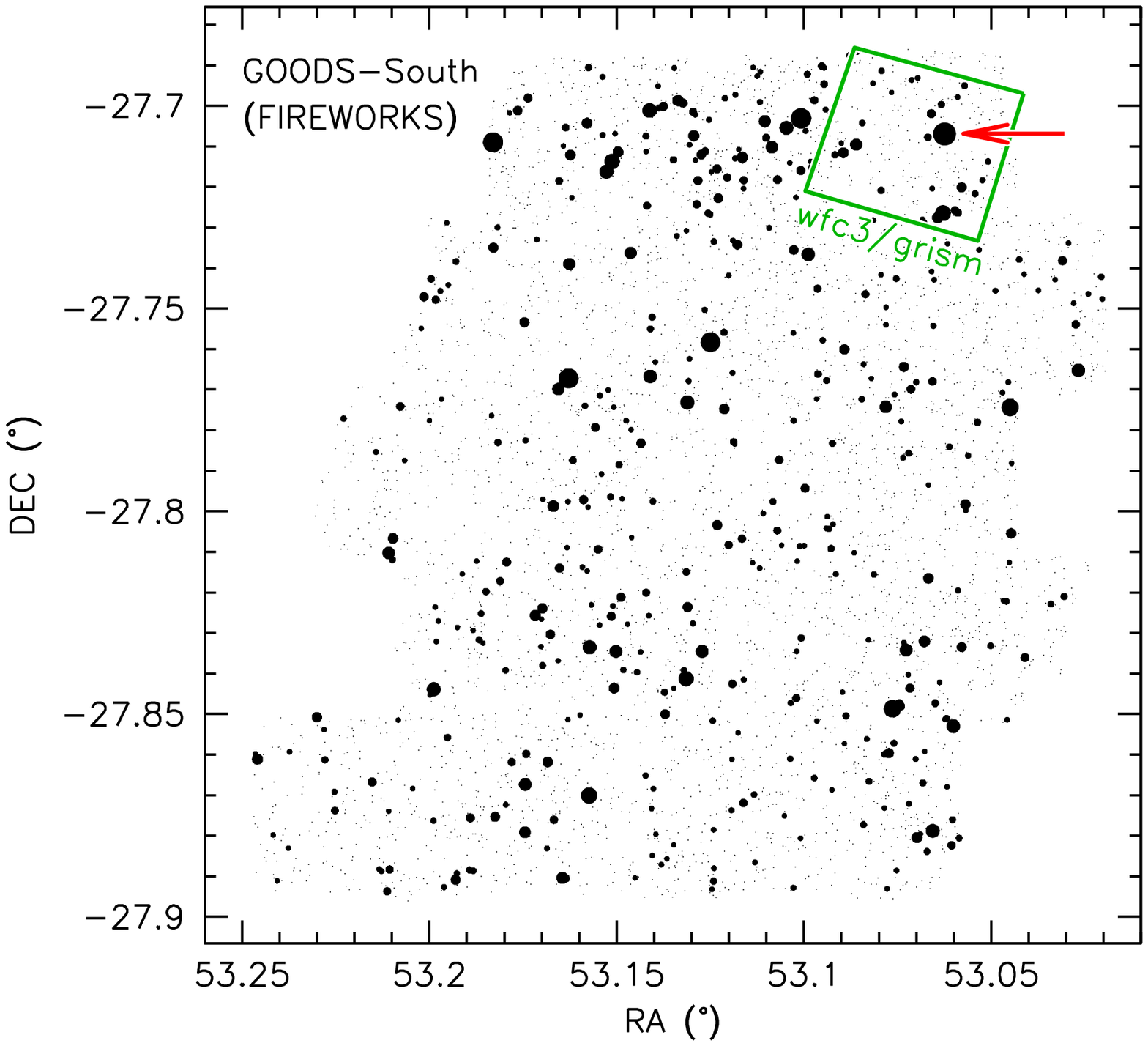}
\caption{\small Galaxies in the GOODS-South field, from the
FIREWORKS catalog (Wuyts et al.\ 2008). Filled circles are galaxies
at $z>1.5$, with the size of the circle indicating the brightness
in the $K$ band. The green box shows the location of the single HST/WFC3
G141 grism exposure that has been obtained as part of the WFC3 ERS.
The arrow indicates the brightest galaxy at $z>1.5$ in this pointing,
object \gala\ in the Wuyts et al.\ catalog.
\label{select.plot}}
\end{figure}
ACS and WFC3 color images of the galaxy are shown in Fig.\ \ref{imspec.plot},
along with the SED
from the Wuyts et al.\ catalog.
The galaxy is faint and unremarkable in the ACS bands but very bright in
the WFC3 images, owing to its red SED. It is composed of a compact core
in addition to diffuse spiral arms, which appear to originate from a
tidal interaction with a companion galaxy,
object \galb\ in the FIREWORKS catalog. The companion
has a similar SED as 4871 but is a factor of $\sim 10$ fainter at
$K=22.0$.
It has a $2\arcsec$ long tidal tail, extending away from \gala.

Interestingly, both \gala\ and \galb\ are X-ray sources
({Luo} {et~al.} 2008, ID numbers 145 and 142 respectively).
Their X-ray luminosities are $6.4\times 10^{43}$\,ergs\,s$^{-1}$
and $3.5\times 10^{43}$\,ergs\,s$^{-1}$ respectively, where we used
the full-band fluxes from {Luo} {et~al.} (2008) and the redshift derived below.
These luminosities would imply star formation rates
$\gg 1000$\,\msun\,yr$^{-1}$ ({Ranalli}, {Comastri}, \&  {Setti} 2003), and we conclude
that both galaxies almost certainly host an active galactic nucleus
(AGN). The AGN in the companion galaxy is likely
heavily obscured: \galb\ has an 8\,$\mu${}m ``upturn'' (see
Fig.\ \ref{imspec.plot}) and is a very bright MIPS 24\,$\mu$m source
with a flux density of $0.4$\,mJy ({Wuyts} {et~al.} 2008).

\gala\ has been targeted several times
for optical spectroscopy.
Three spectroscopic redshifts are available,
all from the GOODS-VIMOS survey: $z=0.352$, $z=2.494$, and $z=2.609$,
with qualities C, C, and B respectively ({Popesso} {et~al.} 2009; {Balestra} {et~al.} 2010).
As we show below all three redshifts are incorrect.

\section{HST WFC3 Grism Spectroscopy}

\noindent
The field was observed with the G102 and G141 grisms, providing
continuous wavelength coverage from 0.8--1.7\,$\mu$m for all
objects in the $2'\times 2'$ WFC3/IR field of view. Each
grism image has a total integration time of 4212\,s, divided over
four dithered exposures in two orbits. We reduced the grism observations
and extracted spectra using a combination of standard {\tt pyraf}
tasks (e.g., {\tt multidrizzle}), the {\tt aXe} package ({K{\"u}mmel} {et~al.} 2009),
and custom scripts to improve background subtraction and optimize
the extraction apertures (see, e.g., {Pirzkal} {et~al.} 2004).
The wavelength calibration and extraction apertures for
G102 and G141 are based on undispersed images in $Y_{098}$
and $H_{140}$ respectively. These direct images were obtained at the
same dither positions as the dispersed data.

The grism spectrum of \gala\ is shown in Fig.\
\ref{spec.plot}; it is of very high
quality with ${\rm S/N}\approx 90$
per 47\,\AA\ pixel at $1.2\,\mu$m. The galaxy has
strong H$\beta$, H$\gamma$, and H$\delta$ absorption lines, and
a  pronounced Balmer break. The redshift $z=1.902\pm 0.002$.
The [O\,{\sc iii}] lines are
undetected; the upper limit on their rest-frame equivalent
width is $\lesssim 2$\,\AA. Note that these lines (and H$\beta$)
are completely inaccessible from the ground, as they fall in between
the $J$ and $H$ atmospheric windows. The average
rest-frame equivalent width of H$\beta$, H$\gamma$,
and H$\delta$ is 8\,\AA\,$\pm 1$\,\AA,
which implies that a post-starburst
population dominates the rest-frame optical light.

We fitted the spectrum with {Bruzual} \& {Charlot} (2003)
stellar population synthesis models (see, e.g., {Kriek} {et~al.} 2009).
Good fits are obtained for populations with low star formation rates
at the epoch of observation but relatively young
luminosity-weighted ages
($\approx 0.5$\,Gyr), combined
with a moderate amount of dust ($A_V \sim 1$). Adopting simple
top-hat star formation histories, we find that the data
can be fit with an extreme burst of $\sim 5000$\,\msun\,yr$^{-1}$
at $z\sim 2.2$ (purple),
or with a star formation rate of $\sim 500$\,\msun\,yr$^{-1}$
sustained over $\sim 1$\,Gyr (orange). In the latter model
the star formation truncated only 150\,Myr prior to the epoch
of observation, comparable to the dynamical time at the distance
of the companion galaxy. Models with less dust and higher
luminosity-weighted ages do not fit the spectrum well
as they have stronger Ca\,H+K and weaker H$\delta$ absorption
than is observed; as an example, the red model in Fig.\ \ref{spec.plot}
has a luminosity-weighted age of 1\,Gyr and $A_V=0.3$ and is a poor
fit to the spectrum.
Scaling the models to the total magnitudes
given in {Wuyts} {et~al.} (2008) we find that the stellar mass of \gala\
is $(4\pm 1)\times 10^{11}$\,\msun\ for a {Chabrier} (2003) IMF.

We also extracted a
spectrum of the companion galaxy from the grism
data, even though it is quite faint at $H=22.4$.
As can be seen in Fig.\ \ref{spec.plot}
we clearly detect the continuum, thanks to the low background
from space and the lack of sky lines.
The galaxy has a continuum break in the same wavelength region as
\gala\ and shows oxygen lines and H$\beta$ in emission.
Its redshift of $z=1.898 \pm 0.003$ is
consistent with that of \gala, demonstrating that the
two are associated. Assuming that the H$\beta$ emission
is due to star formation we derive a star formation rate
of order $5-10$\,\msun\,yr$^{-1}$ 
({Kennicutt} 1998, for $A_V=1-2$\,mag).
Interestingly, the spectrum of the companion
galaxy is redder than that of \gala, although this is difficult
to quantify due to contamination of its spectrum from a nearby
object. This may be caused by dust and/or the presence of
an old stellar population.

\setcounter{figure}{2}
\begin{figure}[htbp]
\epsfxsize=8.5cm
\epsffile{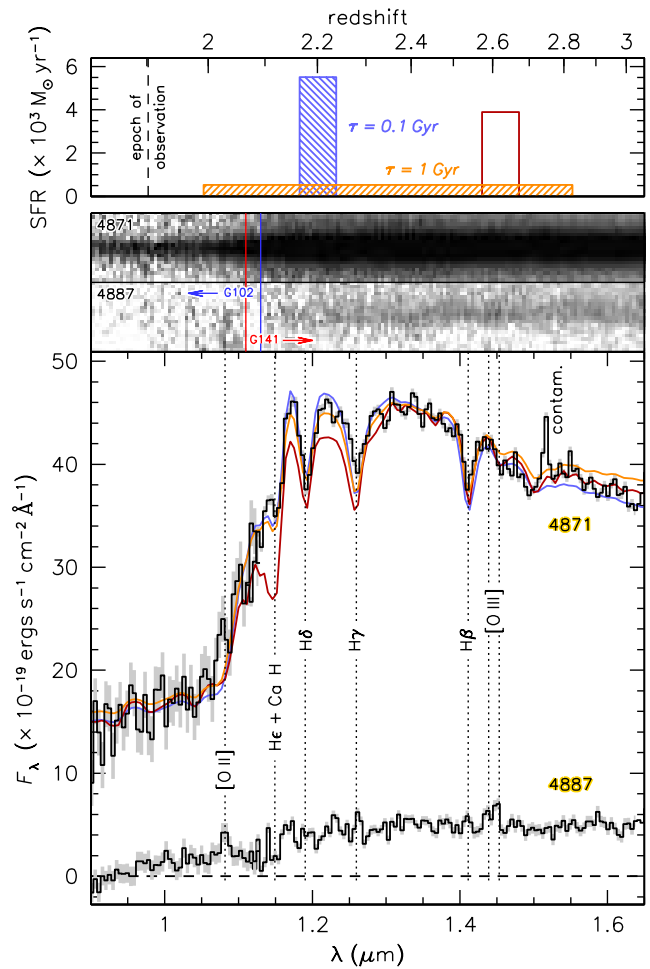}
\noindent
\caption{\small
HST/WFC3 G102 and G141
grism spectra of \gala\ and its companion.
The spectrum of \gala\ is dominated by A stars,
and shows prominent Balmer absorption lines.
The orange and purple lines are stellar populations
with a luminosity-weighted age of $0.5$\,Gyr, $A_V\sim 1$, and a
redshift $z=1.902$. The data are consistent with a short, intense
burst at $z=2.2$ or a more extended star formation history.
The red model has a luminosity-weighted age of 1 Gyr and less
dust; this model can be ruled out.
The feature at 1.51\,$\mu$m is a contaminating
emission line from an unrelated galaxy.
The spectrum of \galb\ also shows a continuum break, along with
emission lines of [O\,{\sc ii}]\,$\lambda 3727$, H$\beta$, and
[O\,{\sc iii}]\,$\lambda 4959,5007$ at approximately
the same redshift
as \gala.
\label{spec.plot}}
\end{figure}

\section{Structure and spatially-resolved spectroscopy}

\begin{figure}[htbp]
\epsfxsize=8.5cm
\epsffile{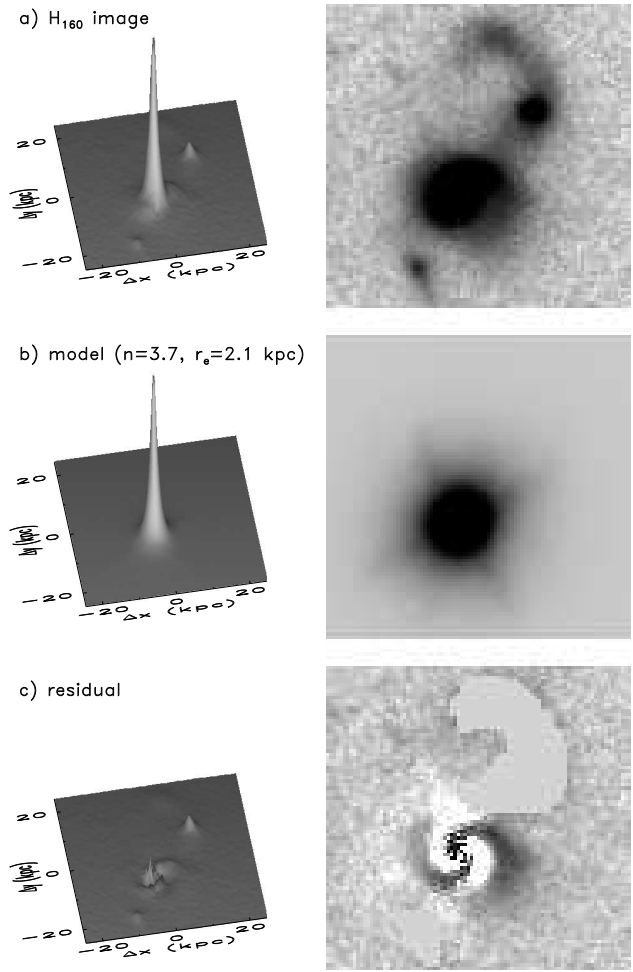}
\caption{\small Sersic fits to the $H_{160}$ image of \gala,
which was drizzled to
a pixel scale of $0\farcs 065$. The galaxy image (a),
the best-fitting model (b), and the residual (c) are shown.
The 3D plots illustrate that most of the light is in a compact
core.  The residual image shows a regular two-armed
spiral, which may have been induced by a tidal interaction.
\label{galfit.plot}}
\end{figure}

\noindent
As discussed in \S\,1,
massive quiescent
galaxies at $z\sim 2$ typically
have very small sizes. Despite its spiral arms
this is also the case for \gala,
as most of its light
comes from a compact core. We quantified this by fitting
{Sersic} (1968) models to the $H_{160}$ image using {\tt galfit}
({Peng} {et~al.} 2002). Other objects in the field,
including the companion galaxy, were masked in the fit.
The fit and the residuals are shown in Fig.\ \ref{galfit.plot}.
The
asymmetric spiral pattern is a striking feature
in the residual image.
The best-fit Sersic index $n=3.7\pm 0.3$
and the best-fit effective radius $r_e=0\farcs 25\pm 0\farcs 03$,
corresponding to $2.1\pm 0.3$\,kpc. The formal
errors are very small; the quoted uncertainties indicate the
full range of solutions obtained when using different stars in the
field as PSFs, but do not include other sources of systematic error.

\noindent
\begin{figure}[htbp]
\epsfxsize=8.5cm
\epsffile{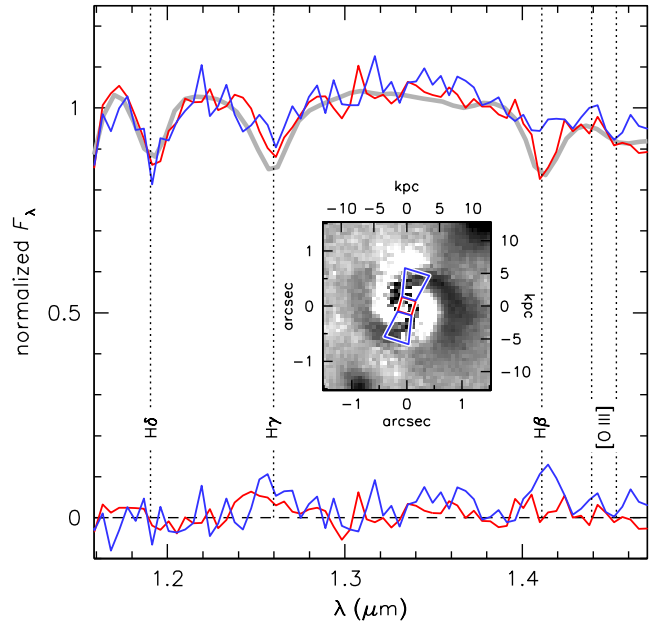}
\caption{\small Spatially-resolved Balmer lines. The red spectrum is
for the central $r<0\farcs 13$ of \gala\ ($r<1$\,kpc)
and the blue spectrum is for radii $0\farcs 13 < r < 0\farcs 65$.
Residual spectra, obtained by subtracting the (light grey) model from
the data, are also shown. At large radii H$\beta$ is filled
in by emission, possibly due to
star formation associated with the spiral arms.
The non-detection of [O\,{\sc iii}] (and [O\,{\sc ii}],
which is not shown) suggests a high metallicity for the gas in
these regions.
\label{2d.plot}}
\end{figure}

The S/N of the grism data is sufficiently high that we can
compare the spectrum of the core to that at larger radii.
As shown in Fig.\ \ref{2d.plot} the average
spectrum of the inner 4 pixels ($r\leq 0\farcs 13$) is similar to
that at large radii ($0\farcs 13<r<0\farcs 65$), with the notable
exception of H$\beta$: it is undetected away from the center, which
implies that it is filled in by emission. We demonstrate this by
subtracting the {Bruzual} \& {Charlot} (2003) model shown in Fig.\ \ref{spec.plot}
from both the central spectrum and the outer spectrum. The spectrum
of the inner parts shows no systematic residuals, but the spectrum away
from the center shows a positive residual at the wavelength of
H$\beta$. We infer that \gala\ is not entirely ``dead'' but is
forming stars in the spiral arms. The amount of star formation
is difficult to quantify and depends on the assumed
reddening; assuming $E(B-V)\sim 0.3$ it is $\sim 20$\,\msun\,yr$^{-1}$.
\vspace{0.5cm}

\noindent

\section{Discussion}

\noindent
The WFC3 grism and imaging data of \gala\ may provide ``smoking gun''
evidence for minor mergers as an important growth mechanism of
massive galaxies: \gala\ is a massive, compact galaxy
at $z\sim 2$ which is interacting with a $\sim 10\times$
less massive companion. The quiescent spectrum of the
primary galaxy is qualitatively consistent with the spectra of
other compact high redshift galaxies and
with the old stellar ages of present-day early-type galaxies.
This mode of growth has been proposed by several recent
studies to explain the size difference between massive
galaxies at high redshift and low redshift
(e.g., {Bezanson} {et~al.} 2009; {Naab} {et~al.} 2009).

Nearby ellipticals have gradients in their color and metallicity,
such that they are bluer and more metal-poor at larger radii
(e.g., {Franx}, {Illingworth}, \&  {Heckman} 1989).
Interestingly, we can begin to address the origin of these
gradients with the kind of data that we are now getting from
HST. The relatively strong oxygen lines and weak H$\beta$ of
the infalling galaxy imply
$\log R_{23}\sim 1$, and a metallicity that is
$\gtrsim 1/3$ times the Solar value ({Pilyugin} \& {Thuan} 2005).
The spectrum extracted from the disk of \gala\ has,
by contrast, no detected oxygen lines and an unambiguous detection
of H$\beta$. It has $\log R_{23}\lesssim 0$, which implies a Solar
or super-Solar metallicity. Qualitatively these
results are consistent with the idea that the metallicity gradients
of elliptical galaxies reflect a gradual increase with radius
in the fraction of stars that came from infalling low-mass satellites.

The apparent absence of star
formation in the central regions of \gala\
might be related to its active nucleus.
It has been suggested by many authors that AGN could prevent gas
cooling and star formation (e.g., {Croton} {et~al.} 2006) and
in this context the observed properties of \gala, such as the
lower limit on the ratio
of its X-ray luminosity to [O\,{\sc iii}] and H$\beta$,
may provide constraints on the mechanism(s) of AGN feedback
(see also {Fiore} {et~al.} 2008; {Kriek} {et~al.} 2007, 2009).
In any case, the fact that both interacting
galaxies host an AGN is remarkable, as it demonstrates that
their black holes are undergoing a ``growth spurt''
prior to their merger.
We note
here that the only indication of the AGNs in the optical and near-IR
is a faint emission line in the VIMOS spectra of
\gala, which we now
identify\footnote{The three erroneous redshifts for \gala\
were not due to a mis-identification of this line; the line was
not recognized as a real feature in the GOODS-VIMOS analysis
of the VIMOS spectrum.} as C\,{\sc iv}. 

There are several important caveats, uncertainties, and complications.
First,  \gala\ is not only growing through the accretion of
\galb, but also through star formation. There is evidence
for star formation in the companion 
(although its emission lines could be influenced by its
active nucleus) and also
in the spiral arms of \gala. In most models such
``residual'' star formation takes place in the center of the
most massive galaxy (see, e.g., {Naab} {et~al.} 2009), but that is in
fact the only place where we do {\em not} see evidence for
star formation. We note, however,
that because of the large mass of \gala\ the
specific star formation rate of the entire system is low at
SFR\,/\,M$_{\rm stellar}$\,$\lesssim 10^{-10}$\,yr$^{-1}$.

Second, although the spectrum of \gala\ resembles those
of the compact  galaxies studied in 
{Kriek} {et~al.} (2006) and {van Dokkum} {et~al.} (2008), the galaxy formed its
stars at significantly lower
redshift. As shown in \S\,3 its star formation rate probably
was $\sim 500$\,\msun\,yr$^{-1}$ as recently as $150$\,Myr
prior to the epoch of observation, i.e., at $z\approx 2$.
It is therefore not a direct descendant of quiescent galaxies at
$z\sim 2.3$. Interestingly,
star forming galaxies at $z>2$ are typically larger
than \gala\ in the rest-frame optical
(e.g., {Toft} {et~al.} 2007), which may imply
that \gala\ is unusual or that a significant fraction of the
star formation in massive galaxies at $z\sim 2.5$
takes place in heavily obscured, compact regions.

Third, the fact that the time since the truncation
of star formation is similar to the dynamical time
calls into question whether we are
witnessing a ``two-stage'' galaxy formation process,
with steady accretion of satellite galaxies following an initial
highly dissipational star formation phase
(e.g., {Naab} {et~al.} 2009; {Dekel} {et~al.} 2009). An alternative interpretation is
that the companion galaxy is somehow related to the truncation,
for example by triggering the AGN in \gala\ $\sim 150$\,Myr ago.
Numerical simulations that aim to reproduce both the 2D spectrum and the
morphological features might shed some light on these issues.


As illustrated in this {\em Letter}
the WFC3 camera on HST has opened up a new regime of detailed
spectroscopic and imaging studies of high redshift galaxies. The quality
of the rest-frame optical continuum spectra
shown in Fig.\ \ref{spec.plot} greatly exceeds what can be achieved
from the ground (see, e.g., Kriek et al.\ 2009), and the grism provides
simultaneous spectroscopy of all 200-300 objects
with $H\lesssim 23$ in the WFC3 field.
Future WFC3 spectroscopic and imaging
surveys over large areas have the potential to robustly
measure the evolution of galaxies over the redshift range $1<z<3$.

\acknowledgements{
We thank the WFC3 ERS team for their exciting program and
Marijn Franx, Hans-Walter Rix, Mariska Kriek, Katherine
Whitaker, and Anna Pasquali for comments.
}



\end{document}